\documentclass{article}
\usepackage{graphicx}

\begin{document}

\begin{center}
{\large HADRONIC CORRELATIONS ABOVE THE CHIRAL/DECONFINEMENT TRANSITION}\\
\medskip
DAVID B.\ BLASCHKE$^{\,1,2}$ and KYRILL A. BUGAEV$^{1,3}$\\
\smallskip
$^1${\em Fachbereich Physik, Universit\"at Rostock, D-18051 Rostock, Germany}\\
$^2${\em Bogolyubov Laboratory for Theoretical Physics, JINR, 141980 
Dubna, Russia}\\
$^3${\em Bogolyubov Institute for Theoretical Physics, Kiev, Ukraine}
\end{center}

\hspace*{-\parindent}  

\begin{abstract}
\noindent
The statistical bootstrap model is critically revised in order to
include a medium-dependent resonance width in it.
We  show that a thermodynamic model with a vanishing width below the 
Hagedorn temperature $T_H$ and a Hagedorn spectrum-like width above 
$T_H$ may not only eliminate the divergence of the thermodynamic
functions above $T_H$, but also gives a satisfactory description
the lattice quantum chromodynamics (QCD) data on the energy density
above the chiral/deconfinement transition as the main result of this 
contribution.
This model allows to explain the absence of heavy resonance 
contributions in the fit of the experimentally measured particle
ratios at SPS and RHIC energies.
\end{abstract}

\bigskip

\hspace*{-\parindent}{\it Keywords}: Statistical bootstrap model, Mott
transition, Spectral function, Lattice QCD thermodynamics

\medskip

\begin{figure}[t]
\leftline{\includegraphics[width=10cm,angle=-90]{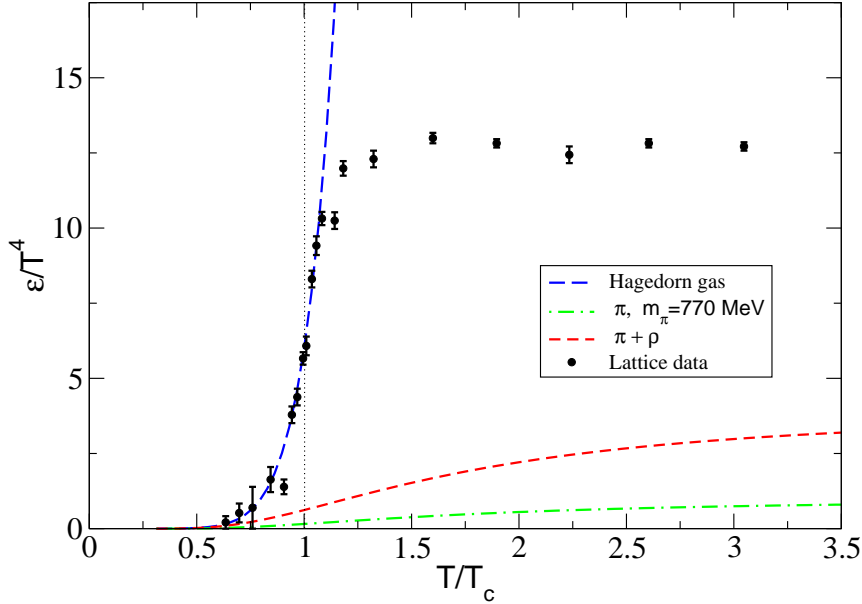}}
\caption{\label{fig1}
Energy density of lattice QCD (circles) and
Hagedorn gas (long dashed line) as functions of the reduced temperature 
$T/T_c$. The contributions of pion gas (dotted line)
and gas of pions and $\rho$ mesons (dashed line)
are shown for a comparison. 
}
\end{figure}

\begin{center} 

{\large\it 1. Introduction}

\end{center}
Ultrarelativistic heavy-ion collision experiments at SPS and
RHIC are performed with the aim to create conditions of temperature and 
density under which hadronic matter undergoes a phase transition to 
the hypothetical quark-gluon plasma (QGP) and to investigate the properties of
this state of matter once it is created.
The strongest theoretical support for the existence of the QGP
comes from lattice QCD simulations at finite temperature $T$ which show a 
step-like enhancement of the effective number of degrees of freedom 
(energy density in units of $T^4$) at a critical temperature $T_c$,
see Fig. \ref{fig1}.  
This behavior is conventionally interpreted as the transition from a few 
mesonic degrees of freedom (mainly $\pi$ and $\rho$ mesons) to those of
quasifree quarks and gluons.
A microscopic description of this transition is, however, still missing.
On the one hand, it has been shown \cite{Tawfik03} that a resonance gas model 
can perfectly explain the steep rise in the number of degrees of freedom at 
$T {\approx} T_c$. 
On the other hand, lattice QCD has also revealed that hadronic correlations 
persist for $T>T_c$ \cite{Petreczky}. 
The question arises whether it is more appropriate to describe hot QCD matter
in terms of hadronic correlations rather than in terms of quarks and gluons.
In the present contribution, we introduce a generalization of 
the Hagedorn resonance gas (statistical bootstrap) model as such a description.

%
The statistical bootstrap model (SBM) \cite{SBM:1,SBM:1B,SBM:2,SBM:3}
is based on the hypothesis that hadrons are made of hadrons,
with constituent  and compound hadrons being treated on the same footing.
This implies an exponentially growing  form of the hadronic mass spectrum
$\rho_H (m) \approx C_H m^{-a} \exp \left[ m/T_H \right]$
for $m \rightarrow \infty$. The parameter $T_H$, Hagedorn temperature, was
interpreted as a limiting temperature reached at infinite energy density.

The extensive investigation of the SBM has led to a formulation of
both the important physical ideas and the mathematical methods for    
modern statistical mechanics of strongly interacting matter.
Thus it has been clarified that the original SBM requires two crucial 
modifications:
\begin{itemize}
\item hadrons should be considered as composite objects (the simplest
way are hadrons as MIT bags \cite{MIT} of quarks and gluons) and, their 
proper volume $v$ has to be taken into
account \cite{Vol:1,Vol:2} in the partition function;
\item only $SU_c(3)$ color singlet clusters of quarks and gluons contribute
\cite{Singl:1,Singl:2,Singl:3,Singl:4}
to the partition function of the system with their masses $m_i$ and volumina 
$v_i$.
\end{itemize}
The  SBM with  proper volume  was solved analytically by the Laplace
transform to the isobaric ensemble \cite{Vol:3} and, indeed, its
solution showed various possibilities for the phase transition between
the QGP and the hadron gas.
Since then this technique has been successfully applied not only to solve  far
more sophisticated versions
\cite{Singl:1,Singl:2,Singl:3,Singl:4,SBM:new}  
of the SBM, but also to find an analytical solution of a simplified version of
the statistical multifragmentation model
\cite{SMM:1,SMM:2} for  the nuclear liquid-gas phase transition.

However, up to now the formulation of the SBM has some severe  problems.
The first one is the absence of a width for the heavy resonances.
From the Particle Data Group \cite{PDG}  we know that heavy
 resonances with masses $m \ge 3.5$ GeV may have width comparable with
their masses. Taking the widths into account shall effectively reduce
the statistical weight of the resonance.  
As we shall show below, this change may eventually remove the
divergence of the SBM thermodynamics.
The second problem arises while discussing the results of the hadron
gas (HG) model \cite{SPS,RHIC}. 
The HG model   describes remarkably well
the light hadron  multiplicities measured in nucleus-nucleus collisions
at CERN SPS \cite{SPS} and BNL RHIC \cite{RHIC} energies.
This model is nothing else as the SBM of light hadrons which accounts for the
proper volume of hadrons with masses below $2.5$ GeV, 
but neglects the  contribution of the exponentially
growing mass spectrum.

In other words, in order to calculate particle ratios within the HG
model it is, on the one hand, necessary to consider all strong  decays of 
resonances according their partial width collected in \cite{PDG}, 
and, on the other hand, it is necessary to truncate the hadron spectrum for  
masses above $2.5$ GeV.
Thus, one immediately faces the following problem:
``Why do the  heavy resonances with masses above $2.5$ GeV predicted by the
SBM not appear in the particle spectra measured in heavy-ion collisions 
at SPS and RHIC energies?''
Note that the absence of heavy resonance contributions in the particle ratios
cannot be due to the statistical suppression of the Hagedorn mass spectrum 
because the latter should not be strong in the quark-hadron  phase transition 
region, where those ratios are believed to be formed \cite{SPS,RHIC}. 

In the present contribution we suggest that the introduction of a finite
width of the resonances can solve the above problems of the SBM.
In the next section we formulate a simple statistical model that incorporates  
besides of the Hagedorn mass spectrum also medium dependent resonance widths
due to the hadronic Mott effect, and analyze its mathematical structure.
In Section 3 we discuss a model fit to recent lattice data of QCD 
thermodynamics \cite{Tawfik03} and some possible consequences for heavy-ion 
physics..  
%

\newpage
\vspace*{0.5cm}

\begin{center}
{\large \it 2.  Resonance Width Model: Mott Transition}
\end{center}
According to QCD, hadrons are not elementary, pointlike objects but rather 
color singlet bound states of quarks and gluons with a finite spatial extension
of their wave function. While at low densities a hadron gas description can
be sufficient, at high densities and temperatures, when hadronic wave functions
overlap, nonvanishing quark exchange matrix elements between hadrons 
occur in order to fulfill the Pauli principle. This leads to a Mott-Anderson 
type delocalization transition with frequent rearrangement processes of color 
strings (string-flip \cite{stringflip}) so that hadronic resonances become 
off-shell with a finite, medium-dependent width. 
Such a Mott transition has been thoroughly discussed for light hadron systems
in \cite{huefner} and has been named {\it soft deconfinement}.
The Mott transition for heavy mesons may serve as the physical mechanism behind
the anomalous J/$\psi$ suppression phenomenon \cite{burau}.

We introduce the width $\Gamma$ of a resonance in the statistical model
with the Hagedorn mass spectrum through  the spectral function 
\begin{equation}
A(s,m)=N_s \frac{\Gamma~m}{(s-m^2)^2+\Gamma^2~m^2}~,
\end{equation} 
a Breit-Wigner distribution of virtual masses with a maximum at  $\sqrt{s}=m$
and the normalization factor
\begin{equation}
\label{two}
%
%
N_s = \left[ \int_{m_b^2}^\infty {ds}
\frac{ \Gamma m }{ ( { s} - m^2)^2 + \Gamma^2 m^2  } \right]^{-1} =
\frac{1}{ \frac{\pi}{2} + 
\arctan \left( \frac{m^2 - m^2_b }{ \Gamma m} \right) }\,.
\end{equation}
The energy density of this model with zero resonance proper volume
for given temperature $T$ and baryonic chemical potential $\mu$
can be cast in the form
\begin{eqnarray}
\varepsilon(T,\mu) &=& 
\sum_{i = \pi, \rho,...} g_i ~\varepsilon_M (T,\mu_i;m_i)\nonumber \\
&& + 
\sum_{A = M,B}  \int_{m_a}^\infty dm \int_{m_b^2}^\infty {ds}
~\rho_H(m)~A(s,m)~\varepsilon_A (T,\mu_A;\sqrt{s}),
\label{one}
\end{eqnarray}
where the energy density per degree of freedom with a mass $m$ is
\begin{equation}
\label{three}
%
%
\varepsilon_A (T,\mu_A;m)  = 
\int  \frac{d^3 k}{ (2 \pi)^3 }  
\frac{\sqrt{k^2+m^2}}{\exp \left(\frac{\sqrt{k^2 +m^2} - \mu_A}{T}
\right)  + \delta_A } \, ,
\end{equation}
with the degeneracy  $g_A$ and the baryonic chemical potential $\mu_A$ 
of hadron $A$. For mesons, $\delta_{M} = -1 ~$,  $\mu_{M} = 0$ and 
for baryons $~ \delta_{B} =  1$~and $\mu_{B}=\mu$, respectively.
According to Eq. (\ref{one})  the energy density of hadrons consists of
the contribution of light hadrons for $m_i < m_A$ ~ and the contribution 
of the Hagedorn mass spectrum $\rho_H(m)$  for $m \ge  m_A$.

A new element of Eq. (\ref{one}) in comparison to the SBM
is the presence of the $ \sqrt{s} $-dependent spectral function.
The analysis shows that, depending on the behavior of the resonance width 
$\Gamma$ in the limit $m \rightarrow \infty$,
there are the following possibilities:
\begin{itemize}
\item For vanishing resonance  width, $\Gamma = 0 $, Eq. (\ref{one})  
evidently reproduces the usual SBM.
\item For  final values of the resonance width, $\Gamma =$ const, 
Eq. (\ref{one}) diverges for all temperatures $T$ because, in contrast to the 
SBM, the statistical factor  in Eq. (\ref{one}) behaves as 
$\left\{\exp \left[ (m_b - \mu_A) / T \right] + \delta_A \right\}^{-1}$
so that it cannot suppress the exponential divergence of the Hagedorn mass 
spectrum $\rho_H (m)$.
\item For a resonance width growing with mass like the Hagedorn spectrum
$\Gamma \sim  C_{\Gamma} \exp \left[ \frac{ m}{T_H} \right] $  or faster,
Eq. (\ref{one}) converges again. 
\end{itemize}
Indeed, in the latter case  the Breit-Wigner spectral function behaves as 
\begin{equation}
\label{four}
%
N_s \frac{ \Gamma m}{ (s - m^2)^2 + \Gamma^2 m^2 }~ 
\biggl|_{m \rightarrow \infty} \rightarrow  
\frac{2}{\pi ~ \Gamma}  \sim  \exp \left( - \frac{m }{ T_H } \right)
\end{equation}
and cancels the exponential divergence of the Hagedorn mass spectrum.
Hence, the energy density remains finite.
Note that both  the analytical properties of model (\ref{one}) and
the right hand side of Eq. (\ref{four}) remain the same, if a Gaussian shape 
of the spectral function is chosen instead of the Breit-Wigner one.

It can be shown that the behavior of the width at finite resonance masses is 
not essential for the convergence of the energy density (\ref{one}). 
In other words, for a convergent energy density (\ref{one})
above $T_H$ it is sufficient to have
a very small probability density (\ref{four}) (or smaller)
for a resonance of mass $m$ to be found in the state with the virtual mass 
$\sqrt{s}$.
Since there is no principal difference between the high and low mass 
resonances, we can use the same functional dependence of the width 
$\Gamma$ for all masses.
Thus, for the following model  ansatz
\begin{equation}
\label{five}
%
%
\Gamma (T) =
\left\{ \begin{array}{ll}
 0\,,  & {\rm for} \hspace*{0.3cm} T \le T_H  \,,  \\
 & \\
C_{\Gamma}~ \left( \frac{ m}{T_H} \right)^{N_m}
\left( \frac{ T}{T_H} \right)^{N_T} \exp \left( \frac{ m}{T } \right)\,,
& {\rm for } \hspace*{0.3cm} T >  T_H  \,,
\end{array} \right.
\end{equation}
the energy density (\ref{one}) is finite for all temperatures and 
the divergence of the SBM is removed.
At $T = T_H$, depending on choice of parameters,  it may have 
either a discontinuity or its partial $T$ derivative may be discontinuous.
As discussed above, for $T \le T_H$ such a model corresponds to the usual SBM, 
but for high temperatures
$T > T_H$ it remains finite for a wide choice of powers $N_m$.

\newpage
\vspace{0.5cm}

\begin{center}
{\large\it 3. Applications for lattice QCD and heavy-ion collisions}
\end{center}

\begin{figure}[t]
\leftline{\includegraphics[width=10cm,angle=-90]{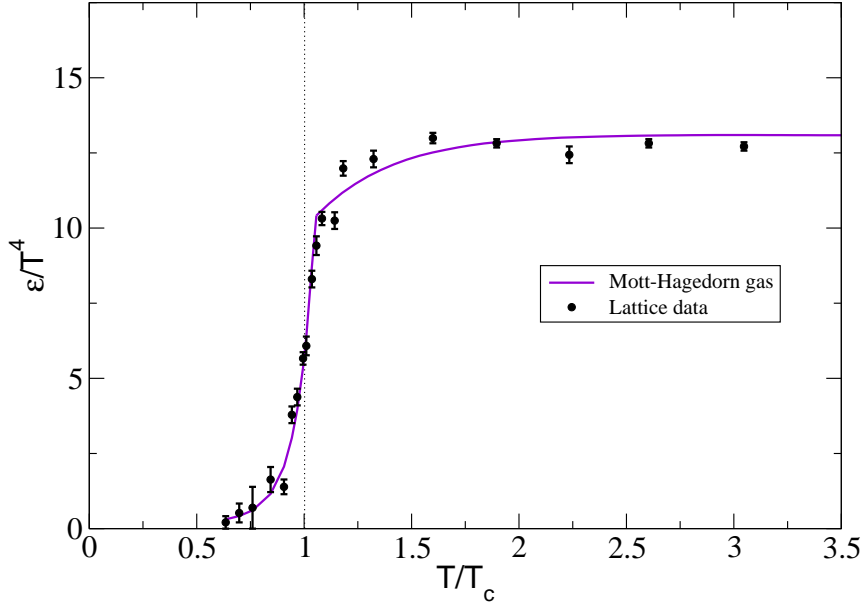}}
\caption{\label{fig2}
Fit of the lattice QCD data \cite{Tawfik03} with the Mott-Hagedorn resonance 
gas model (\ref{five}). For details see text.
}
\end{figure}

As one can see from Fig. \ref{fig1} the Hagedorn gas model correctly 
reproduces the
lattice QCD results below the critical temperature $T_c$ and just in a
vicinity above $T_c$, but not for large temperatures.  
Fig. \ref{fig2} shows a comparison of the same lattice QCD data  
\cite{Tawfik03}
with the Mott-Hagedorn gas (\ref{five}) where the parameters of the spectral 
function are  $N_T=2.325$, $N_m = 2.5$ and $T_H=165$ MeV and $m_a=m_b=1$ GeV. 
The successful description of the lattice energy density \cite{Tawfik03} 
indicates that above $T_c$ the strongly interacting matter may be well 
described in terms of strongly correlated hadronic degrees of freedom.
This result is based on the concept of soft deconfinement and provides an 
alternative to the conventional explanation of the deconfinement transition
as the emergence of quasifree quarks and gluons.
%

Another interesting  feature of the model (\ref{five}) is 
that it allows to explain naturally the absence of
heavy resonance contributions to the particle yields measured at
highest SPS and all RHIC energies, where QGP conditions are expected 
\cite{SPS,RHIC}.
In order to find out whether a given resonance has a chance to survive till 
the freeze-out it is necessary to compare its lifetime with the typical 
timescale in the system.
There are two typical timescales usually discussed in nucleus-nucleus 
collisions, the equilibration time $\tau_{eq}$ and the formation time $\tau_f$.
The equilibration time tells when the matter created in collision process
reaches a thermal equilibrium which allows to use the hydrodynamic and 
thermodynamic descriptions.
For Au + Au collisions at RHIC energies it was estimated to be about 
$\tau_{eq} \approx 0.5$ fm \cite{Taueq}.  
On the other hand in transport calculations  the formation time is used:
the time for constituent quarks to form a hadron.  
The formation time depends on the momentum and energy of the created hadron,
but is  of the same order $\tau_f \approx 1 - 2$ fm \cite{Tauf} as the 
equilibration time.

Since within our model the QGP is equivalent to a resonance gas with medium
dependent widths, all hadronic resonances with life time $\Gamma^{-1} (m)$ 
shorter  than  $\max\{\tau_f, \tau_{eq}\} $
will have no chance to be formed in the system. 
Therefore, the upper limit of the the integrals over the resonance mass $m$ 
and over the virtual mass $\sqrt{s}$ in  Eq. (\ref{one})
should be reduced to a resonance mass defined by 
\begin{equation}
\label{six}
%
%
\Gamma (m)^{-1} =  \max\{\tau_f, \tau_{eq}\} \,.
\end{equation}
\noindent
This reduction may essentially weaken the energy density gap at the transition
temperature  or even make it vanish.
Thus, the explicit time dependence should be introduced into
the resonance width model (\ref{one}) while applying it to nuclear collisions, 
and this finite time (size) effect, as we discussed, may change essentially
the thermodynamics of the hadron resonances formed in the nucleus-nucleus 
collisions.

\begin{center}
{\large\it 4. Conclusions and Remarks}
\end{center}

The statistical bootstrap model allows to interpret the QGP as the hadron 
resonance gas dominated by the state of infinite mass (and infinite volume).
We argue  that it is necessary to include the resonance width into the SBM 
in order to avoid the contradiction with the experimental data on hadron 
spectroscopy.
We found that the simple model (\ref{one})-(\ref{five}) with a 
vanishing width below Hagedorn temperature $T_H$
and a Hagedorn spectrum-like width above $T_H$ may not only eliminate the 
divergence of the thermodynamic functions above $T_H$, but it is able to 
successfully describe the lattice QCD data \cite{Tawfik03}
for energy density with three fitting parameters only.
Such a model also allows to naturally 
explain the absence of heavy resonance contributions in the
fit of the experimentally measured particle ratios at SPS and RHIC energies.

However, such a modification of the SBM requires an essential change in our 
view on QGP: it is conceivable that hadrons of very large masses which should 
be associated with a QGP cannot be formed
in nucleus-nucleus collisions because of their very short lifetime.  

It is also necessary to remind that presented model should be applied to 
analyze the experimental data with care: it can be successfully applied  
to describe either the quantities associated with the chemical freeze-out, 
i.e. particle ratios or spectra of $\Omega$ hyperons, $\phi$, $J/\psi$ and 
$\psi^\prime$ mesons that are freezing out 
at hadronization \cite{BUGGG:02, GBG, BD, TLS}. 
But as discussed in Refs.
\cite{BUG:96, BUG:99a, BUG:99b} the model  presented  here   should not 
be used for the post freeze-out momentum spectra of other hadrons
produced in the nucleus-nucleus collisions. 
Perhaps only such weakly interacting hadrons like $\Omega$, $\phi$, $J/\psi$ 
and $\psi^\prime$ will allow us to test the model presented. 

The question how to derive the non-zero width below $T_H$ has to be 
investigated.
To solve this problem it will be necessary to include into
present  model a non-zero proper volume of hadrons.  
This, however, will require to consider a mixture of hadrons of different sizes
as studied recently in \cite{MULT:1,MULT:2} or even a relativistic 
modification of the excluded volume of hadrons \cite{MULT:2,RELVDW}.

\begin{center}
\textit{Acknowledgments}
\end{center}
We are grateful to M. I. Gorenstein for interesting discussions and
important comments. K.A.B. acknowledges the financial support of DFG grant 
No. 436 UKR 17/13/03 and the Ministery for Culture and Education of 
Mecklenburg-Western Pomerania.

\medskip





\end{document}